\documentclass[12pt,fleqn]{article}

\usepackage[cp1251]{inputenc}
\usepackage{epsf}

\usepackage[centertags]{amsmath}
\usepackage{amsfonts}
\usepackage{amssymb}
\usepackage{amsthm}

\begin{document}

\baselineskip19pt
\begin{center}
{\bf {\Large  "Partial" quantum cloning and quantum cloning of the
mixed states}} \vspace{0.3cm}

{\bf A.Ya.Kazakov } \vspace{0.5cm} \\Laboratory of quantum information, \\%
St.-Petersburg state university of aerospace instrumentation,

67 B.Morskaya Str., St.-Petersburg, 190000 Russia \vspace{0.2cm}

\vspace{0.5cm}
\end{center}

\begin{abstract}
We discuss the "partial" quantum cloning of the pure two-partite
states, when the "part" of initial state related to the one qubit
is copied only. The same approach gives the possibility to design
the quantum copying machine for the  mixed qubit states.

\end{abstract}

PACS: 03.67 -a, 03.67 Dd

\section{Introduction}
Main laws of quantum mechanics forbids the perfect cloning of the
quantum states, see corresponding discussion for the pure states
in \cite{Wooters}, \cite{Dieks}, and for the mixed states in
\cite{Caves}. But it is possible to carry out an approximate
copying of the quantum states \cite{Buzek}. Quantum cloning
machines (QCM) depend on the conditions accepted at its designing.
They can produce identical copies of the initial state (symmetric
QCM), nonidentical copies (non-symmetric QCM), the quality of the
copying can be either identical for all states (universal QCM) or
depend on the state (state-dependente QCM). Detailed discussion of
the different variants of QCM and theirs possible applications in
quantum cryptography and quantum informatics can be found in
 \cite{Gisin}, \cite{Cerf}.

One possible application of the QCM is an eavesdropping of the
quantum channel. The aim of such eavesdropping defines the main
properties of the designing QCM. One can design QCM which copies
only part of the quantum state, for instance. Such QCM can be
useful if eavesdropper, usually called Eve, intends to catch part
of the transmitted quantum information only. Some classical
analogue of this situation can be classical eavesdropping of the
key words in the transmitted classical information. At quantum
cloning we can choose the different parts of the quantum signal in
which we are interested. In this paper we intend to discuss some
"partial" QCM, which copies one constituent of the two-partite
states.

Our approach gives the possibility to consider QCM for a mixed
states too. It is well known fact, that any mixed state can be
considered as a reduction of a pure state, which is called
"purification" of the mixed state \cite{Preskill}. So, cloning of
the mixed state can be considered as a "partial" cloning of the
"purification" of the mixed state. Some difference between the
"partial" cloning machine and the cloning machine for the mixed
states is connected with the corresponding difference of the sets
of the initial states, see details below. Note, that the main
attention in the present literature was devoted to the cloning of
the pure states  \cite{Gisin}, \cite{Cerf}.

\section{"Partial" quantum cloning machine}
We consider two-partite qubit states, qubits are elements of
two-dimensional Hilbert space ${\bf H}$. In order to construct QCM
we need in tensor product of three such spaces on the ancilla
space:  ${\bf H}_1\otimes {\bf H}_2\otimes {\bf H}_3\otimes {\bf
H}_4 $, here different components are marked by indexes. The first
and third qubit components constitute a quantum state which
carries information in the quantum channel, and the state of first
component is interesting for Eve. The second component is a blank
state, where we will copy the first component, the last component
is necessary for the realization of the QCM.  Let quantum channel
carries the quantum state $\mid \Psi >\in {\bf H}_1\otimes {\bf
H}_3,$
\begin{equation}
 \mid \Psi
>=a_{00}\mid 0_10_3>+a_{01}\mid 0_11_3>+a_{10}\mid
1_10_3>+a_{11}\mid 1_11_3>, \label{init}
\end{equation}
where normalization condition holds,
\begin{equation}
\mid a_{00}\mid ^2+\mid a_{01}\mid ^2+\mid a_{10}\mid ^2+\mid
a_{11}\mid ^2=1.  \label{nor}
\end{equation}
Here and below $\mid 0>,\mid 1>$ are base vectors  in $\bf{H}$. We
suppose, that Eve's goal is a copying of the first component of
this state. After tracing one can obtain:
\begin{equation}
\rho _{init}=Tr_3\mid \Psi ><\Psi \mid =A\mid 0><0\mid +B\mid 0><1\mid +%
\overline{B}\mid 1><0\mid +C\mid 1><1\mid , \label{red}
\end{equation}
\[
A=\mid a_{00}\mid ^2+\mid a_{01}\mid ^2,B=a_{00}\overline{a_{10}}+a_{01}%
\overline{a_{11}},C=\mid a_{10}\mid ^2+\mid a_{11}\mid ^2=1-A.
\]
So, Eve has to realize the cloning to produce the pair of states
(in the first and second components respectively) closest to $\rho
_{init}.$ We consider here symmetric QCM, so, we suppose, that
states in the first and second components have to coincide. Then
produced state must be symmetric with regard to permutation of the
first and second components. Let us introduce the orthonormal
basis in the subspace of ${\bf H_1 \otimes H_2}$ symmetric
regarding this permutation:
\[
\mid \Phi _1>=\mid 0_10_2>,\mid \Phi _2>=\frac 1{\sqrt{2}}\left(
\mid 1_10_2>+\mid 0_11_2>\right) ,\mid \Phi _3>=\mid 1_11_2>.
\]
Let's assume, that the second component be in state $\mid 0
>$ initially. Description of the QCM is, in essence, the definition
of the corresponding unitary operator $U$. Following to
\cite{Buzek}, \cite{Choudhury}, we set
\begin{equation}
 U\mid 0_10_20_4>=\mid \Phi _1>\mid Q_0>+\mid \Phi _2>\mid
Y_0>. \label{CCM1}
\end{equation}
\begin{equation}
U\mid 1_10_20_4>=\mid \Phi _3>\mid Q_1>+\mid \Phi _2>\mid Y_1>,
\label{CCM2}
\end{equation}
where $\mid Q_0>, \mid Q_1>,\mid Y_0>,\mid Y_1>$ are some vectors,
belonging to ${\bf H_4}$. Symmetry of QCM is provided by the fact,
that right-hand part of this relation contains linear combinations
of vectors $\mid \Phi _k> $  only. Taking into account
(\ref{init}), we obtain:
\[
\mid \Xi >=U\mid \Psi 0_20_4>=
\]
\[
\mid \Phi _1>\left( a_{00}\mid 0_3>+a_{01}\mid 1_3>\right) \mid
Q_0>+\mid \Phi _2>\left( a_{00}\mid 0_3>+a_{01}\mid 1_3>\right)
\mid Y_0>+
\]
\[
\mid \Phi _3>\left( a_{10}\mid 0_3>+a_{11}\mid 1_3>\right) \mid
Q_1>+\mid \Phi _2>\left( a_{10}\mid 0_3>+a_{11}\mid 1_3>\right)
\mid Y_1>.
\]
Generally speaking, the choice of the unitary operator $U$ is very
broad and corresponding analysis is quite complex even for the
lowest dimensions, so usually one admits some additional
restrictions. We suppose as in  \cite{Choudhury}, that following
conditions (which guarantee the unitarity of $U$) are fulfilled :
\begin{equation}
<Q_k\mid Q_k>+<Y_k\mid Y_k>=1,k=1,2,  \label{norm1}
\end{equation}
\begin{equation}
<Y_0\mid Y_1>=<Q_0\mid Q_1>=<Q_k\mid Y_k>=0,k=1,2.  \label{norm2}
\end{equation}
Let
\begin{equation}
<Q_0\mid Q_0>= <Q_1\mid Q_1>=\zeta , <Y_1\mid Q_0>= <Q_1\mid
Y_0>=\nu \sqrt{(1-\zeta )\zeta }, \label{con1}
\end{equation}
 so as  $<Y_0\mid Y_0>=<Y_1\mid Y_1>=1-\zeta $, $0\leq \mid \nu \mid \leq 1
$. In this case QCM produces the next state from ${\bf H_1 \otimes
H_2}$:
\begin{equation}
\rho _{out}^{(12)}=Tr_{34}\mid \Xi ><\Xi \mid = \label{out12}
\end{equation}
\[
A\zeta \mid \Phi _1><\Phi _1\mid  +(1-A)\zeta \mid \Phi _3><\Phi
_3\mid +(1-\zeta ) \mid \Phi _2><\Phi _2\mid +
\]
\[
+B\nu \sqrt{\zeta (1-\zeta )}(\mid \Phi _1><\Phi _2\mid + \mid
\Phi _2><\Phi _3\mid ) + \overline{B\nu }\sqrt{\zeta (1-\zeta
)}(\mid \Phi _2><\Phi _1\mid +\mid \Phi _3><\Phi _2\mid ) .
\]
Reducing this state on the first component, we obtain:
\begin{equation}
\rho _{out}^{(1)}=Tr_{234}\mid \Xi ><\Xi \mid =\widetilde{A}\mid 0><0\mid +%
\widetilde{B}\mid 0><1\mid +\overline{\widetilde{B}}\mid 1><0\mid +%
\widetilde{C}\mid 1><1\mid , \label{out}
\end{equation}
\[
\widetilde{A}=1/2-\zeta( 1/2-A),\widetilde{C}=1/2+\zeta(1/2-A)
,\widetilde{B}=B\nu\sqrt{2\zeta (1-\zeta )}.
\]
It is necessary to compare the initial state and state which is
produced by the QCM, in other words, we have to choose the measure
of the closeness of these states. There are different measures,
specifically, fidelity. It is defined for the mixed states as
$F=\left[ Tr\sqrt{\sqrt{\rho _{init}}\rho _{out}^{(1)}\sqrt{\rho _{init}}}%
\right] ^2$, this value is not very suitable for the analytical
considerations. We use here more convenient measure:
\[
\parallel \rho _{init}^{(1)}\otimes \rho _{init}^{(2)}-\rho _{out}^{(12)}\parallel
^2=Tr\left[ \rho _{init}^{(1)}\otimes \rho _{init}^{(2)}-\rho
_{out}^{(12)}\right] ^2 =W(\zeta ,\nu , \Psi ),
\]
where
\[
W(\zeta ,\nu ,\Psi )=A^2(A-\zeta )^2+(1-A)^2(1-A-\zeta
)^2+(1-\zeta )^2+2\left[ A^2(1-A)^2+2\mid B\mid ^4\right] +
\]
\[
2\mid B\mid ^2\left( \mid \sqrt{2}A-\nu \sqrt{(1-\zeta )\zeta } \mid ^2+\mid \sqrt{2}%
(1-A)-\nu \sqrt{(1-\zeta )\zeta }\mid ^2\right) -
\]
\[
2(1-\zeta )\left[ A(1-A)+\mid B\mid ^2\right] .
\]
 This value estimates the difference between initial and final states
 with fixed parameters $a_{00}, a_{01}, a_{10}, a_{11}$. For the determination
 of the QCM parameters we average this value respect to the set of all initial
 states.   We use here the next parametrization of the initial state  $\mid \Psi
>$:
\[
a_{00}=\cos \theta _1,a_{01}=\exp (i\gamma _1)\sin \theta _1\cos
\theta _2,
\]
\[a_{10}=\exp (i\gamma _2)\sin \theta _1\sin \theta _2\cos \theta
_3,a_{11}=\exp (i\gamma _3)\sin \theta _1\sin \theta _2\sin \theta
_3 ,
\]
where
\[
0\leq \theta _k\leq \pi /2,0\leq \gamma _m\leq 2\pi .
\]
Here the first component has zero phase due to the corresponding
freedom of the choice. For the averaging we need in corresponding
measure. Supposing that all states $\mid \Psi >$ are equiprobable,
we choose as such a measure
\begin{equation}
G(\zeta ,\nu )=\parallel \rho _{init}^{(1)}\otimes \rho
_{init}^{(2)}-\rho _{out}^{(12)}
\parallel _{aver}^2=  \label{integ1}
\end{equation}
\[
\frac{1}{\pi ^5}\int_0^{\pi /2}d\theta _1\int_0^{\pi /2}d\theta
_2\int_0^{\pi /2}d\theta _3\int_0^{2\pi }d\gamma _1 \int_0^{2\pi
}d\gamma _2\int_0^{2\pi }d\gamma _3\sin ^2\theta _1\sin \theta
_2W(\zeta ,\nu , \Psi )
\]
Simple calculations lead to the conclusion, that $G(\zeta ,\nu )$
takes its minimal value at $\nu =1$, $\zeta \approx 0.725$. This
value $\nu $ implies, that vectors  $Q_0,Y_1$ and  $Q_1,Y_0$ are
parallel. The values of the fidelity
 $F=\left[ Tr\sqrt{\sqrt{\rho _{init}}\rho _{out}^{(1)}\sqrt{\rho _{init}}}%
\right] ^2$, calculated at $\zeta =0.725$, for the states on the
"real" part of the Bloch sphere,
\[
\rho =\frac 12\left(
\begin{array}{cc}
1+r\cos \theta  & r\sin \theta  \\
r\sin \theta  & 1-r\cos \theta
\end{array}
\right),
\]
$0\leq r \leq 1$, $0\leq \theta \leq \pi /2$, are plotted in
fig.1.  Evidently, that this QCM is state-dependent one, because
the quality of the cloning depends on the quantum state.

\section{QCM for the mixed states.}
The construction described above can be used for the cloning of
the mixed states. Note, that final state produced by QCM depend on
the reduction of the initial state $\rho _{init}$ (\ref{red})
only. It means, that one can reverse our considerations and take
the mixed state $\rho _{init}$ as initial one. Then pure state
$\mid \Psi >$ defined by relation (\ref{init}) belongs to the
space of the larger dimension and it is a "purification" of the
state  $\rho _{init}$. "Purification" of the given mixed state
$\rho _{init}$ can be realized by different methods, as it follows
from (\ref{init}), but this nonuniqueness has not influence in the
results.  QCM constructed in accordance with relations
(\ref{CCM1}), (\ref{CCM2}) produces states  (\ref{out12}),
(\ref{out}), which depend on the parameters of the initial state
$\rho _{init}$ only. But the set of the initial states is
changing, one has to use another parametrization for this set. As
such parametrization one can take a Bloch sphere, see
\cite{Preskill}. Namely, density matrix $\rho _{init}$ can be
described as
\[
\rho _{init}=\left(
\begin{array}{cc}
A & B \\
\overline{B} & C
\end{array}
\right) =\frac 12\left(
\begin{array}{cc}
1+P_3 & P_1-iP_2 \\
P_1+iP_2 & 1-P_3
\end{array}
\right))
\]
where
\[
P_1^2+P_2^2+P_3^2\leq 1.
\]
In the spherical coordinates we have:
\[
P_3=r\cos \theta ,P_1=r\sin \theta \cos \varphi ,P_2=r\sin \theta
\sin \varphi ,
\]
\[
0<r\leq 1,0\leq \theta \leq \pi ,0\leq \varphi \leq 2\pi .
\]
and
\[
A=\frac 12(1+r\cos \theta ),B=\frac 12r\sin \theta \exp \left(
-i\varphi \right) ,C=\frac 12(1-r\cos \theta ).
\]
In order to obtain the parameters of the QCM one has to average
the value $W(\zeta ,\nu ,\rho _{init} )$ on the Bloch sphere. We
suppose that all states in the Bloch sphere are equiprobable, so
averaging is reduced to the integral
\[
G(\zeta ,\nu )=\frac{3}{4\pi }\int_0^1r^2dr\int_0^\pi \sin \theta
d\theta \int_0^{2\pi }d\varphi W(\zeta ,\nu ,\rho _{init} ).
\]
Note, that if Eve has a priori information about transmitted
quantum information she has to choose corresponding weight
multiplier. We are searching in such values $ \zeta ,\nu $ which
correspond to the minimal value $ G(\zeta ,\nu )$. As a result we
obtain  $\nu =1, \zeta \approx 0.715$. The plot of fidelity for
these parameters differs in a small way from the preceding one, so
we omit it.
\section{Conclusion}
We have discussed here a "partial" quantum cloning, when only one
component of the two-partite pure state is cloning. Such cloning
can be considered as a  variant of the eavesdropping of the
quantum channel. The choice of the parameters of the QCM was
realized with help of the some natural criterion. Namely, we seek
in parameters  corresponding to the minimum of the integral
average of the "distance" between the initial state and output
state. Note, that fidelity of the initial and output states for
the most part of the Bloch sphere exceeds value 5/6 which
corresponds to the universal QCM for the pure states \cite{Buzek}.
This fact has two reasons. Firstly, described QCM copies only part
of the two-partite state. Second, this QCM is state-dependent one,
and this non-universality raises the quality of the cloning.

Moreover, here was discussed the cloning of the mixed qubit
states. In order to consider such device we use the "purification"
of the mixed state and then apply "partial" cloning machine. Let's
emphasize, that "purification" of the mixed state is not unique,
but the output state produced by our QCM does not depend on this
nonuniqueness. Parameters of the QCM was sought by minimization of
the integral average of the distance between the initial and
output states. Note, that this averaging differs from the used
above because the sets of the states are different. In this case
the fidelity of the initial and output states on the real part of
the Bloch sphere exceeds the value 5/6 too. Evidently, that choice
of the parameters $\nu ,\zeta $ for the QCM is defined by the
strategy of the eavesdropping.

W have discussed here only one possible QCM for the mixed states.
Evidently, that there are many other variants for the QCM, may be,
without restrictions like (\ref{CCM1}), (\ref{CCM2}),
(\ref{con1}), asymmetric QCM etc.

\begin{figure}[t]
\leavevmode \centering{\epsfbox{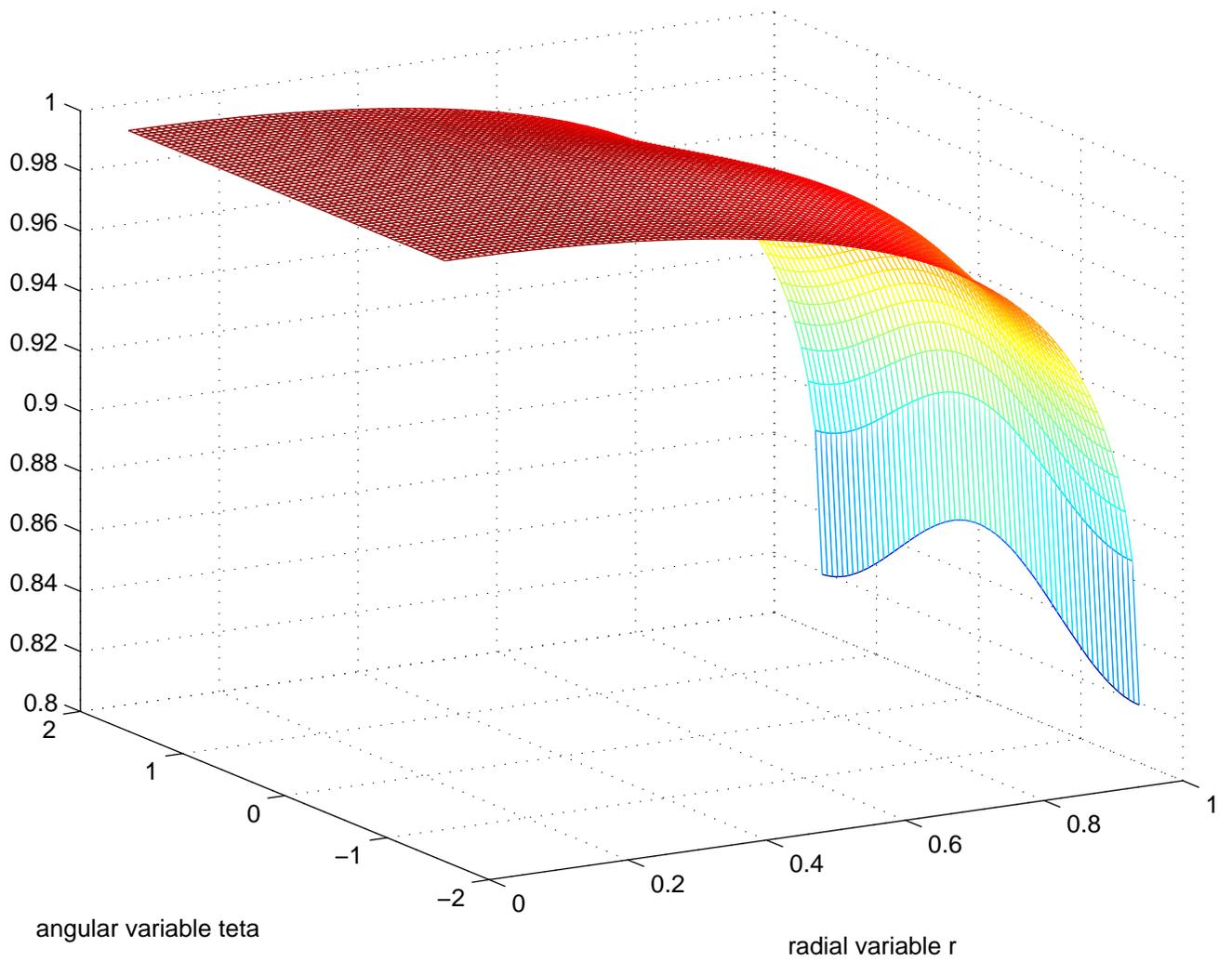}} \caption[]{Fidelity on
real part of Bloch sphere}
\end{figure}

\end{document}